\documentclass[aps,prb,twocolumn,showkeys,groupedaddress,superscriptaddress,nobibnotes]{revtex4}

\usepackage{amssymb}
\usepackage{graphicx}
\usepackage{dcolumn}
\usepackage{bm}
\usepackage{color}
\usepackage{subfig}
\usepackage[normalem]{ulem}
\usepackage[none]{hyphenat}
\usepackage{siunitx}
\usepackage{soul}

\newcommand{\MSbT}{MnSb$_2$Te$_4$}
\newcommand{\MBT}{MnBi$_2$Te$_4$}

\newcommand{\BS}{Bi$_2$Se$_3$}

\newcommand{\BT}{Bi$_2$Te$_3$}

\newcommand{\mnbi}{Mn$_\text{Bi}$\,}
\newcommand{\tebi}{Te$_\text{Bi}$\,}

\newcommand{\bite}{Bi$_\text{Te}$\,}
\newcommand{\bimn}{Bi$_\text{Mn}$\,}

\newcommand{\mnte}{Mn$_\text{Te}$\,}
\newcommand{\PreserveBackslash}[1]{\let\temp=\\#1\let\\=\temp}

\begin{document}

\title[Native point defects and Dirac point gap at \MBT(0001)]{Native point defects and their implications for the Dirac point gap at \MBT(0001)}

\author{M. Garnica}
\email{manuela.garnica@imdea.org}
\affiliation{Instituto Madrileño de Estudios Avanzados en Nanociencia (IMDEA-Nanociencia), 28049 Madrid, Spain}

\author{M.\,M. Otrokov}
\email{mikhail.otrokov@gmail.com}
\affiliation{Centro de F\'{i}sica de Materiales (CFM-MPC), Centro Mixto CSIC-UPV/EHU,  20018 Donostia-San Sebasti\'{a}n, Basque Country, Spain}
\affiliation{IKERBASQUE, Basque Foundation for Science, 48011 Bilbao, Spain}

\author{P. Casado Aguilar}
\affiliation{Instituto Madrileño de Estudios Avanzados en Nanociencia (IMDEA-Nanociencia), 28049 Madrid, Spain}
\affiliation{Departamento de F\'{\i}sica de la Materia Condensada,
Universidad Aut\'{o}noma de Madrid, 28049 Madrid, Spain}

\author{I.\,I.~Klimovskikh}
\affiliation{Saint Petersburg State University, 198504 Saint Petersburg, Russia}

\author{D.~Estyunin}
\affiliation{Saint Petersburg State University, 198504 Saint Petersburg, Russia}

\author{Z.S.~Aliev}
\affiliation{Azerbaijan State Oil and Industry University, AZ1010 Baku, Azerbaijan}
\affiliation{Institute of Physics, National Academy of Sciences of Azerbaijan, AZ1143  Baku, Azerbaijan}

\author{I. R. Amiraslanov}
\affiliation{Institute of Physics, National Academy of Sciences of Azerbaijan, AZ1143  Baku, Azerbaijan}
\affiliation{Baku State University, AZ1148 Baku, Azerbaijan}

\author{N. A. Abdullayev}
\affiliation{Institute of Physics, National Academy of Sciences of Azerbaijan, AZ1143  Baku, Azerbaijan}
\affiliation{Baku State University, AZ1148 Baku, Azerbaijan}

\author{V. N. Zverev}
\affiliation{Institute of Solid State Physics, Russian Academy of Sciences, Chernogolovka, Russia}

\author{M. B. Babanly}
\affiliation{Institute of Catalysis and Inorganic Chemistry, Azerbaijan National Academy of Science, AZ1143 Baku, Azerbaijan}
\affiliation{Baku State University, AZ1148 Baku, Azerbaijan}

\author{N. T. Mamedov}
\affiliation{Institute of Physics, National Academy of Sciences of Azerbaijan, AZ1143  Baku, Azerbaijan}

\author{A.\,M.~Shikin}
\affiliation{Saint Petersburg State University, 198504 Saint Petersburg, Russia}

\author{A. Arnau}
\affiliation{Centro de F\'{i}sica de Materiales (CFM-MPC), Centro Mixto CSIC-UPV/EHU,  20018 Donostia-San Sebasti\'{a}n, Basque Country, Spain}
\affiliation{Departamento de F\'{\i}sica de Materiales UPV/EHU, 20080 Donostia-San Sebasti\'{a}n, Basque Country, Spain}
\affiliation{Donostia International Physics Center (DIPC), 20018 Donostia-San Sebasti\'{a}n, Basque Country, Spain}

\author{A. L. V\'azquez de Parga}
\affiliation{Instituto Madrileño de Estudios Avanzados en Nanociencia (IMDEA-Nanociencia), 28049 Madrid, Spain}
\affiliation{Departamento de F\'{\i}sica de la Materia Condensada,
Universidad Aut\'{o}noma de Madrid, 28049 Madrid, Spain}
\affiliation{Instituto  "Nicol\'{a}s Cabrera", Universidad
Aut\'{o}noma de Madrid, 28049 Madrid, Spain}
\affiliation{Condensed Matter Physics Center (IFIMAC),
Universidad Aut\'{o}noma de Madrid, 28049 Madrid, Spain}

\author{E.\,V. Chulkov}
\email{evguenivladimirovich.tchoulkov@ehu.eus}
\affiliation{Departamento de F\'{\i}sica de Materiales UPV/EHU, 20080 Donostia-San Sebasti\'{a}n, Basque Country, Spain}
\affiliation{Donostia International Physics Center (DIPC), 20018 Donostia-San Sebasti\'{a}n, Basque Country, Spain}
\affiliation{Saint Petersburg State University, 198504 Saint Petersburg, Russia}
\affiliation{Tomsk State University, 634050 Tomsk, Russia}

\author{R. Miranda}
\email{rodolfo.miranda@imdea.org}
\affiliation{Instituto Madrileño de Estudios Avanzados en Nanociencia (IMDEA-Nanociencia), 28049 Madrid, Spain}
\affiliation{Departamento de F\'{\i}sica de la Materia Condensada,
Universidad Aut\'{o}noma de Madrid, 28049 Madrid, Spain}
\affiliation{Instituto  "Nicol\'{a}s Cabrera", Universidad
Aut\'{o}noma de Madrid, 28049 Madrid, Spain}
\affiliation{Condensed Matter Physics Center (IFIMAC),
Universidad Aut\'{o}noma de Madrid, 28049 Madrid, Spain}

\date{\today}

\begin{abstract}
The Dirac point gap at the surface of the antiferromagnetic topological insulator \MBT\, is a highly debated issue. While the early photoemission measurements reported on large gaps in agreement with theoretical predictions, other experiments found vanishingly small splitting of the \MBT\, Dirac cone.
Here, we study the crystalline and electronic structure of \MBT(0001) using scanning tunneling microscopy/spectroscopy (STM/S), micro($\mu$)-laser angle resolved photoemission spectroscopy (ARPES), and density functional theory (DFT) calculations. Our topographic STM images clearly reveal features corresponding to point defects in the surface Te and subsurface Bi layers that we identify with the aid of STM simulations as \bite antisites (Bi atoms at the Te sites) and \mnbi substitutions (Mn atoms at the Bi sites), respectively. X-ray diffraction (XRD) experiments further evidence the presence of cation (Mn-Bi) intermixing. Altogether, this affects the distribution of the Mn atoms, which, inevitably, leads to a deviation of the \MBT\, magnetic structure from that predicted for the ideal crystal structure. Our transport measurements suggest that the degree of this deviation varies from sample to sample. 
Consistently, the ARPES/STS experiments reveal that the Dirac point gap of the topological surface state is different for different samples/sample cleavages. 
Our DFT surface electronic structure calculations show that, due to the predominant localization of the topological surface state near the Bi layers, \mnbi\, defects can cause a strong reduction of the \MBT\, Dirac point gap, given the recently proved antiparallel alignment of the \mnbi moments with respect to those of the Mn layer. Our results provide a key to puzzle out the \MBT\, Dirac point gap mystery.
\end{abstract}

\vspace{2pc}

%
\maketitle
%
%

\section{Introduction}

\begin{figure}[!bth]
	\begin{center}
		\includegraphics[width=0.48\textwidth]{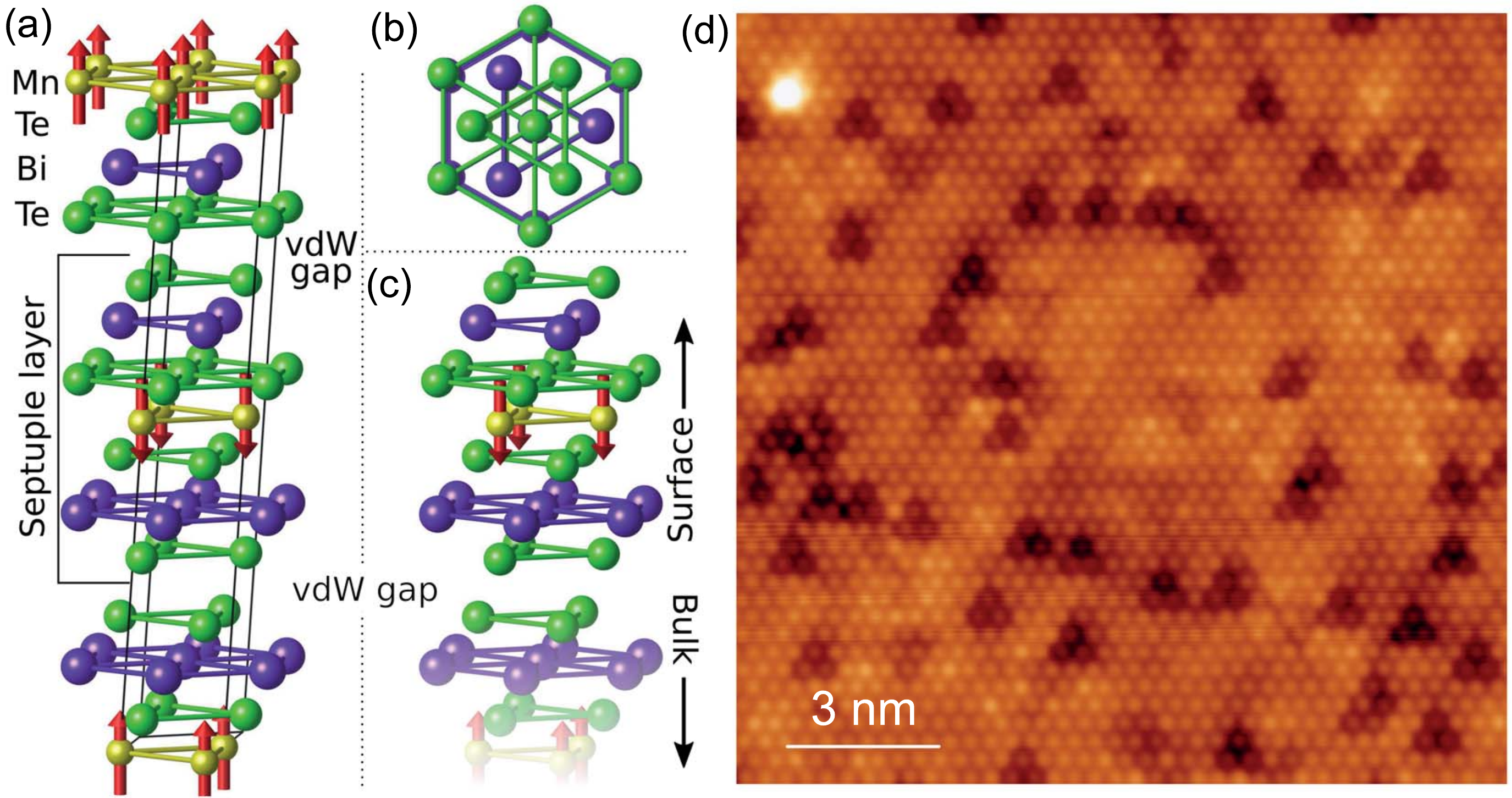}
	\end{center}
	\caption{(a) Side view of the bulk crystal structure of \MBT\, with red arrows showing the interlayer AFM order. The crystal cleavage in this block-layered compound takes place at the van der Waals gap, thus exposing a Te layer of a SL at the surface. (b) Top and (c) side views of the surface crystal structure. (d) Atomically resolved STM image of the surface of the cleaved sample ($1\,$V and $0.3\,$nA) showing dark triangular depressions and a bright circular protrusion.} \label{fig:str}
\end{figure}

The interplay between magnetism and topology is a fertile ground for new exotic ground states in condensed matter \cite{Tokura.nrp2019}. In this context, intrinsic magnetic topological insulators (TIs) have attracted a great deal of attention \cite{Otrokov.2dmat2017, Otrokov.jetpl2017, Eremeev.jac2017, Hirahara.nl2017, Hagmann.njp2017,  Otrokov.nat2019, Otrokov.prl2019, Li.sciadv2019, Zhang.prl2019, Gong.cpl2019, Lee.prr2019, Yan.prm2019,  Vidal.prb2019, Yan.prb2019, Rienks.nat2019, Chen.ncomms2019, Wu.sciadv2019, Li.prl2020, Hu.ncomms2020, Klimovskikh.npjqmat2020, Hirahara.ncomms2020, Wimmer.advmat2021, Eremeev.jpcl2021, Gao.nat2021} due to the recent discovery of the first representative of this class, i.e., the van der Waals antiferromagnetic (AFM) compound \MBT\, \cite{Otrokov.nat2019, Otrokov.prl2019, Li.sciadv2019, Zhang.prl2019, Gong.cpl2019}. 
This material crystallizes in the trigonal $R\bar 3m$-group structure \cite{Lee.cec2013, Aliev.jac2019, Zeugner.cm2019}, made of septuple layer (SL) blocks, in which atomic layers are stacked in the Te-Bi-Te-Mn-Te-Bi-Te sequence, as shown in Fig.\ref{fig:str}a. Neighboring SLs are bound by van der Waals forces. Below 25 K, \MBT\, orders antiferromagnetically due to the antiparallel alignment between alternate, ferromagnetically-ordered Mn layers \cite{Otrokov.nat2019, Yan.prm2019, Li.arxiv2020-2}, with the local moments pointing out-of-plane (Fig. \ref{fig:str}a). The combination of these crystalline and magnetic structures makes \MBT\, invariant with respect to the $S=\Theta T_{1/2}$-symmetry (where $\Theta$ is time-reversal and  $T_{1/2}$ is primitive-lattice translation), which gives rise to the $Z_2$ topological classification of AFM insulators\cite{Mong.prb2010,Fang.prb2013} ($Z_2=1$ for this material \cite{Otrokov.nat2019, Li.sciadv2019, Zhang.prl2019}). According to the bulk-boundary correspondence principle, the topological surface state appears in the bulk bandgap of a TI, which in case of the AFM TI might be gapped at the $S$-breaking crystal termination\cite{Mong.prb2010,Fang.prb2013}. For \MBT, the $S$-breaking surface is (0001), which, according to \emph{ab initio} calculations, is indeed gapped due to the uncompensated out-of-plane FM Mn layer \cite{Otrokov.nat2019, Zhang.prl2019, Li.sciadv2019}. A plethora of exotic phenomena can be hosted if the Fermi level of the experimentally synthesized samples lies inside the Dirac point (DP) gap, such as various kinds of the quantized Hall effect \cite{Deng.sci2020, Ge.nsr2020, Liu.ncomms2021, Deng.nphys2021, Mong.prb2010}, axion insulator states \cite {Liu.nmat2020, Hu.sciadv2020, Eremeev.jpcl2021}, Majorana fermions \cite{Peng.prb2019}, chiral hinge modes \cite{Perez-Piskunow.arxiv2021}, etc. 

The experimental studies of the  \MBT\, surface electronic structure have reported contradictory results, with some groups finding a gapped Dirac cone (gap of at least 60 meV and larger) \cite{Otrokov.nat2019, Lee.prr2019, Vidal.prb2019, Estyunin.aplmat2020, Zeugner.cm2019}, in agreement with theoretical predictions \cite{Otrokov.nat2019, Zhang.prl2019, Li.sciadv2019}, while others revealing a gapless topological surface state \cite{Hao.prx2019, Li.prx2019, Chen.prx2019, Swatek.prb2020, Nevola.prl2020, Yan.arxiv2021}.
A recent photoemission study reports a reduced DP gap of about 20 meV in some \MBT\, samples \cite{Shikin.srep2020}. Given the complex crystal structure of \MBT, the problem may lie in whether the bulk and/or surface of real samples faithfully reproduce the predicted properties of the ideal crystal structure (shown in Figs. \ref{fig:str}a-c), in particular, the magnetic ones. This is especially important in view of the possible applications since the envisaged quantum devices \cite{Zhang2012,An.npjcmat2021} would often employ few SL-layer thick films. 

In this paper, we report on a combined study of the AFM TI \MBT (0001)\, surface using low temperature scanning tunneling microscopy/spectroscopy (STM/S), high-resolution micro($\mu$)-laser angle resolved photoemission spectroscopy (ARPES), and density functional theory (DFT) calculations. High-resolution STM images complemented by the STM simulations allow us to observe, identify and provide a detailed characterization of two types of point defects: the \bite antisites (i.e., Bi atoms at the Te sites) located in the surface layer and \mnbi substitutions (Mn atoms at the Bi sites) in the second atomic layer. The fingerprints of these defects appear as circular protrusions and triangular depressions, respectively, and are readily seen in the topographic images at relatively large bias voltages (Fig.  \ref{fig:str}d).
The presence of the \mnbi substitutions in the second layer strongly suggests that they should also occur in the sixth (Bi) layer as well, while Bi atoms, in turn, should occupy Mn positions in the fourth layer (\bimn). This is indeed confirmed by our structure refinement of the X-ray diffraction data.

Importantly, since the defects in the second, fourth, and sixth layers essentially involve Mn, they cause deviations of the magnetic structure from the ideal one (shown in Figs. \ref{fig:str}a-c) to a ferrimagnetic\cite{Riberolles.prb2021, Liu.prx2021, Lai.prb2021}, which might influence the DP gap size. In line with this, our STS measurements reveal that, depending on the sample cleavage, the local density of states is compatible with both large ($\sim$50 meV) and small ($<$20 meV) DP gaps, in agreement with the laser-ARPES experiments, detecting that the DP gap changes from sample to sample. Our DFT surface electronic structure calculations show that the \mnbi\, defects cause a strong reduction of the \MBT\, DP gap due to the antiparallel alignment of the \mnbi moments with respect to those of the Mn layer\cite{Riberolles.prb2021, Liu.prx2021, Lai.prb2021} and predominant localization of the topological surface state near the Bi layers.
We thus attribute the variation of the DP gap in the same sample (as observed by STS) or different samples (ARPES) to a different degree of the defectness of the \MBT\, crystals at local or global structure level, respectively.
This is also supported by the results of our transport measurements that reveal a variation of the N\'eel temperature in a series of the \MBT\, single crystal samples. Our results are instrumental in unifying seemingly contradictory reports concerning the \MBT\, DP gap and stress a necessity of suppressing the cation (Mn-Bi) intermixing thus reducing the number of the \mnbi\, defects in this AFM TI.

\section{Results}

\begin{figure*}[!bth]
	\begin{center}
	\includegraphics[width=\textwidth]{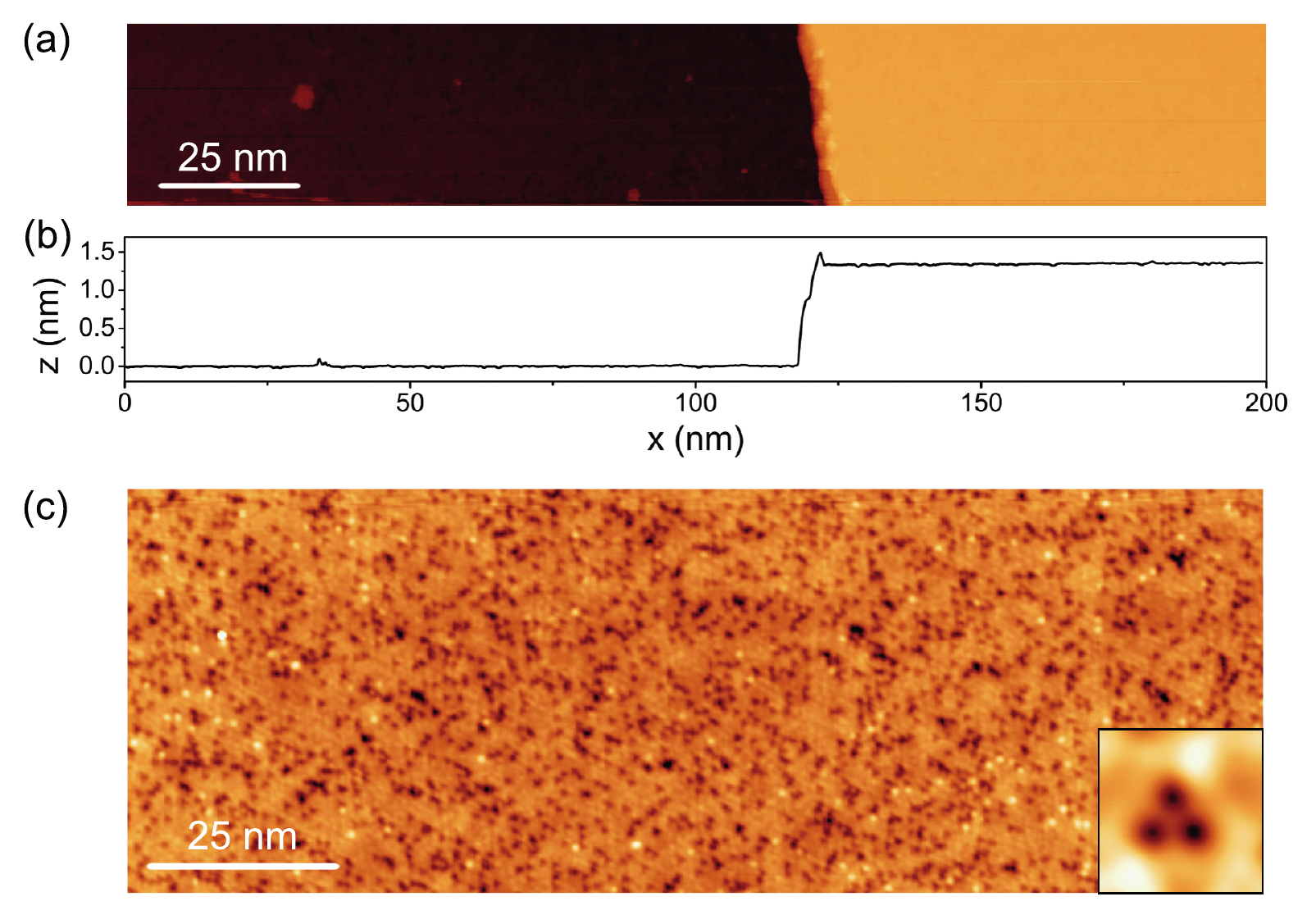}
	\end{center}
	\caption{Large scale STM topographic images of the \MBT\, Te-terminated (0001) surface after \textit{in-situ} cleavage showing (a) flat terraces separated by a SL thick step ($2\,$V and $0.5\,$nA)  and (b) its corresponding line profile. (c) STM image of a representative area far from the step where the characteristic density of surface defects can be observed ($1\,$V and $0.1\,$nA). Inset: Zoom of $2.5$nm$^2$ of one of the  triangular-shaped defects taken at $1\,$V and $0.1\,$nA.} \label{STM}
\end{figure*}

Figure \ref{fig:str}d shows an atomically-resolved STM image of the \MBT\, crystal (0001) surface after cleavage in ultra high vacuum. A hexagonal lattice with a lateral periodicity of $4.28\pm0.05\,$\AA\ is resolved in agreement with the bulk $a$ lattice constant measured by X-ray diffraction ($a$=4.33 \AA, see Supplementary Information as well as Ref. \cite{Aliev.jac2019}). Since the (0001) plane is the natural cleavage plane containing van der Waals bonded Te layers, the surface is terminated by the outmost Te layer of a SL. On a large scale, see Fig. \ref{STM}, the surface shows atomically flat terraces several hundreds of nm in size. They are separated by steps with a height of about $13.7 \pm0.5\,$\AA\, (Figs. \ref{STM}a,b), in good agreement with the expected value of the thickness of a single \MBT\, SL, which is roughly equal to one third of the hexagonal $c$ parameter, i.e., $40.93$\, \AA\, (see Supplementary Information). The topographic STM image of the \MBT\, flat terrace reveals randomly distributed triangular defects with an average density of the order of $3.4 - 5.2\% $  (Fig. \ref{STM}c and its inset), similar to those at the Te-terminated transition metal dichalcogenides surfaces \cite{Zhussupbekov2021}, as well as bright atomic-size protrusions, although much less abundant.

\begin{figure*}[!bth]
	\begin{center}
		\includegraphics[width=0.7\textwidth]{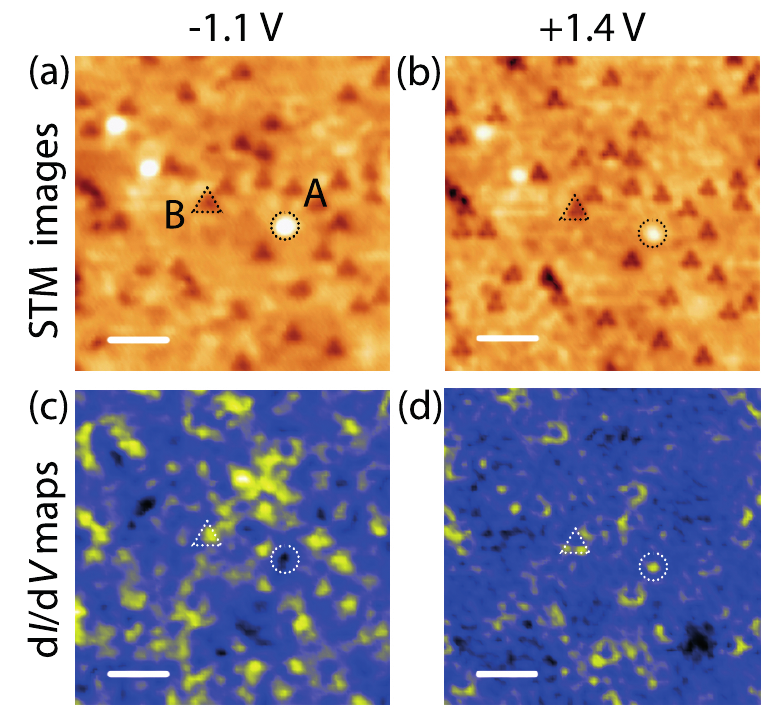}
	\end{center}
\caption{Topographic STM images taken at $1.2\,$K on the same area showing (a) the occupied ($-1.1\,$V) and (b) unoccupied ($1.4\,$V) states and their corresponding d$I$/d$V$ maps (c) and (d), respectively. We observe two different types of defects located in the surface Te layer (type A) and in the subsurface Bi layer (Type B). The scale bar is 2 nm. For the bias-dependent topographs and d$I$/d$V$ maps in the [-1.5 V : +1.5 V] range see Supplementary Fig. S2.} \label{defects}
\end{figure*}

Atomically-resolved images and their corresponding d$I$/d$V$ maps taken at negative (occupied states) and positive (unoccupied states) sample bias voltages $V$ confirm the presence of two types of point defects (labeled as A and B, Fig.~\ref{defects}), randomly distributed over the \MBT(0001) surface. The type B defect, marked with a small dashed triangle, is the most abundant and shows a bias dependent appearance \cite{note_contrast}. At large biases, both negative (Fig.~\ref{defects}a) and positive (Fig.~\ref{defects}b), these defects appear dark and are especially pronounced. 
However, towards the low bias voltages, e.g. near the Fermi level or in the energy region within the bulk band gap where the topological surface state is located, these features loose their well-defined triangular shape (Supplementary Figs. S2, S3 and S4). Most clearly, their three-fold symmetry is resolved for $V>0$, as shown in Fig. \ref{defects}b and inset of Fig. \ref{STM}c, with the three dark spots located at the positions of surface Te atoms (see also Fig. \ref{fig:str}d and Supplementary Figs. S2, S5). Furthermore, the fact that the separation between the three dark spots in a single triangular-shaped depression corresponds to the \MBT(0001) lateral lattice constant $a$ indicates that the defects causing these features are located in the subsurface (Bi) layer. A variation of bias voltage between -1.7 eV and +2.4 eV reveals that the appearance of all dark triangular defects evolves in the same fashion (see Supplementary Fig. S2 for the part of these data in the [-1.5 V : +1.5 V] range). A careful counting allows estimating their concentration in the range $3.4 - 5.2\% $ of the Bi sites, depending on the cleavage or surface area (see Supplementary Figure S6). To get deeper insight into the nature of the type B defects, we have performed STM simulations using DFT (see Supplementary Note VI and Methods section). As seen in Supplementary Fig. S7, the simulated topographic images are consistent with type B defects being \mnbi. Moreover,
among all hypothetically possible defects in the Bi layer, i.e., Bi vacancy, \tebi antisite or \mnbi substitution, the latter has the lowest formation energy \cite{Hou.acsn2020, Du.advfunmat2020}.
To further verify the point defect behind the triangular-shaped feature, we resort to the X-ray diffraction measurements. They permit to identify a cation disorder in the Mn and Bi positions (Supplementary Fig. S1 and the corresponding note). The structure refinement performed yields the amount of Mn atoms at Bi sites (\mnbi) of about 4.6\%, which is in a reasonable agreement with the concentration of second-layer defects seen in our STM measurements as well as results of other X-ray diffraction experiments \cite{Zeugner.cm2019}. Moreover, recent neutron diffraction measurements reported in Refs. \cite{Riberolles.prb2021, Ding.prb2020-2}  detect \mnbi atoms, too. Thus, we attribute the triangular-shaped depressions to the \mnbi\, substitutions in the subsurface layer. Similar features have been previously observed in STM for magnetically doped \BS-family TIs \cite{Hor.prb2010, Lee.pnas2015, Zhang2017} and, recently, for \MBT(0001) \cite{Yuan.nl2020,Yan.prm2019,Liang.prb2020}. Besides, the existence of  \mnbi\, defects has also been claimed recently based on the electron energy loss spectroscopy and transmission electron microscopy analysis \cite{Hou.acsn2020}.

Another atomic scale defect clearly observed in the STM topographs is a bright circular protrusion, referred to as type A. They are well seen only at relatively large bias voltages, i.e., $V \gtrsim |0.9|$ eV (see Fig. \ref{defects}a,b and Supplementary Figs. S2 and S5). Superimposing a 2D hexagonal lattice on the topograph with atomic resolution yields the lateral location of the defect coinciding with that of the surface Te atoms sites (Fig. \ref{fig:str}d). This is consistent with the circular shape of the features, suggesting that they are either incorporated in the surface Te layer or adsorbed on top of it. The bright appearance excludes the possibility of them being Te vacancies, which are usually resolved as depressions at Te-terminated surfaces \cite{Netsou.acsn2020, Zhussupbekov2021}. The small measured apparent height of $0.5$ \AA\, at $1\,$V as well as the difficulty to manipulate it with the STM tip points towards the substitutional character of this defect.  
According to recent calculations \cite{Hou.acsn2020, Du.advfunmat2020}, the lowest formation energy for atomic defects in the van der Waals Te layers in bulk corresponds to \bite\, antisites (Bi atoms substituting the Te atoms), while the formation energies of the \mnte antisite or Te vacancy are much larger. Our STM simulations support the hypothesis of the type A defect being \bite, since the feature's appearance as a bright protrusion is reproduced for both occupied and unoccupied states (Supplementary Fig. S7 and the corresponding note). Therefore the type A defects can be identified with the \bite\, antisites, similarly to what happens in \BT\, TI \cite{Netsou.acsn2020}, where they also appear as bright protrusions for both bias polarities. 
This conclusion is in line with previous STM studies of \MBT(0001) \cite{Yuan.nl2020,Yan.prm2019,Liang.prb2020, Huang.prm2020}. On the basis of our measurements, \bite appear less frequently than \mnbi (triangular depressions), with an estimated concentration ranging between 0.02 and $0.35$\,\% of the Te layer (depending on the surface location). Such a small concentration does not allow their reliable detection using XRD.

Thus, based on the acquired STM topographic images, we can solidly claim the presence of point defects in the two topmost atomic layers of the \MBT(0001) surface.
For the first-layer defect \bite (type A), d$I$/d$V$ maps show a clear change of contrast when going from -1.1 eV (occupied states, Fig.\ref{defects}c), where it appears dark, to 1.4 eV (unoccupied states, Fig.\ref{defects}d), where it looks bright. The same behaviour, but inverted, is observed for the \mnbi (type B) defects. Our LDOS simulations reveal the change of contrast for both \mnbi and \bite (see Supplementary Fig. S8 and the corresponding note), which further supports the defects assignment.

In an attempt to find signatures of the defects lying below the second layer we focus in the areas where neither first- nor second-layer defects are observed, at least in abundance. Interestingly, even  though topographic images do not show any special feature as can be seen in Supplementary Fig. S3, we have observed the appearance of extended bright features d$I$/d$V$ maps at $-0.4$ V and $-0.15$ V.  
Their size appears to be approximately equal to nine or six in-plane lattice parameters $a$, respectively. 
Although based on the STM data it is hardly possible to deduce to which layer the corresponding defects belong, the extension of these features points towards a relatively deep location of the defects. 

As it has been mentioned above, our structure refinement based on the XRD measurements indicate the existence of the cation (Mn-Bi) intermixing, whose signatures are clearly seen in STM as well. The latter means that, apart from the second (subsurface) layer, \mnbi atoms should also occur in the sixth layer counting from the surface, while Bi atoms, in turn, should occupy Mn positions (\bimn) in the fourth layer. However, no clear signatures of the defects lying below the second atomic layer have been observed on the STM topographic images so far\cite{Yuan.nl2020,Yan.prm2019,Liang.prb2020, Huang.prm2020}. As far as the spectroscopic d$I$/d$V$ imaging is concerned, apart from the above discussed deep-lying defects (Supplementary Fig. S3), a feature with a lateral size of 3$a$ has been observed on d$I$/d$V$ maps taken at about -0.08 V and attributed to \bimn \cite{Huang.prm2020}. We do not observe such a feature in our d$I$/d$V$ measurements. Nevertheless, as shown in Supplementary Information, the presence of \bimn atoms is indeed confirmed by our structure refinement. 
The results of our structure characterizations are in agreement with the previous X-ray \cite{Zeugner.cm2019} and neutron \cite{Riberolles.prb2021, Ding.prb2020-2} diffraction studies, as well as with the conclusions based on electron energy loss spectroscopy and transmission electron microscopy \cite{Lee.prr2019, Hou.acsn2020}.
It is not surprising that the signatures of \bimn and \mnbi lying in the fourth and sixth layers are not clearly seen on topographies as the tunneling probability depends drastically of the tip-sample distance. In addition, the corresponding features (whose extension should be about several lattice parameters as a minimum) may laterally overlap with each other, making them hardly distinguishable.
Nevertheless, based on the agreement between the concentrations of the \mnbi substitutions measured by STM and XRD one can conclude that these and \bimn appear already in bulk before the crystal cleavage. 

\begin{figure*}[!bth]
\begin{center}
\includegraphics[width=1\textwidth]{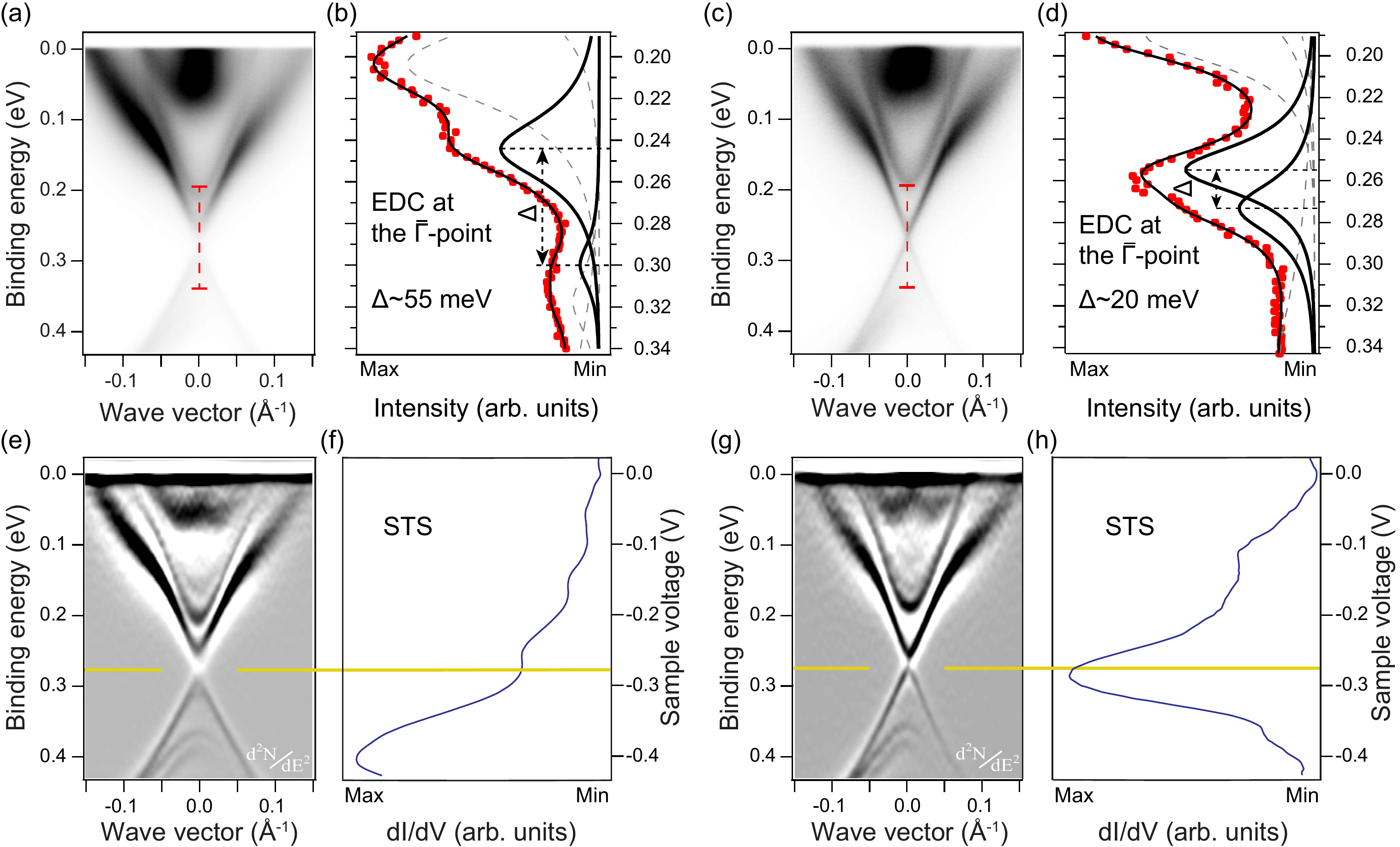}
\end{center}
\caption{(a,c) Measured \MBT(0001) ARPES dispersions corresponding to a larger (a) and smaller (c) DP gaps (measurements conditions: photon energy $h\nu=6.3$ eV; temperature $T=10$ K). (b,d) Measured (red points) and fitted (solid black curves) $\overline{\Gamma}$-point EDCs acquired at binding energies close to the gapped DP with decomposition on the spectral components shown. The peaks corresponding to the upper and lower Dirac cone parts are shown by the bold dashed black lines, while those of the bulk conduction band -- by the thin dashed grey lines. The binding energy intervals of the presented EDCs correspond to the intervals marked in (a,c) by vertical red lines. 
EDCs fitting yields the DP gap values of $\Delta \sim 55$ meV (b) and $\Delta \sim 20$ meV (d). (e,g) Second-derivative ($\text{d}^2N/\text{d}E^2$) representation of the data shown in (a,c), respectively, providing a better visualization of the larger (e) and smaller (g) DP gaps. (f,h) Spatially averaged tunneling conductance spectra showing a clear dip (f) and peak (h) at the expected energy position of the gapped DP. The spectra shown in (f,h) are compatible with larger ($\sim$50 meV) and smaller ($<$20 meV) DP gaps, respectively. The horizontal yellow lines show the correspondence between the ARPES and STS spectra. The d$I$/d$V$ curves in (f,h) are shown in wider bias voltage range in Supplementary Figs. S4,S5, respectively. The ARPES data in (a) and (c) correspond to two different samples, while the STS data in (f) and (g) have been acquired from the yet another sample, but after different cleavages.} \label{ARPES}
\end{figure*}

Let us now discuss the surface electronic structure of our \MBT\, samples based on the results of the laser-ARPES and STS measurements performed in the AFM state (Fig. \ref{ARPES}). In ARPES, the linearly dispersing topological surface state (TSS) is clearly visible in Fig. \ref{ARPES}a. The energy distribution curve (EDC) at the $\overline \Gamma$ point (red dotted curve in Fig. \ref{ARPES}b) presents a minimum at binding energies of about $0.27$ eV, indicating the presence of a gap at the DP. According to the EDC fitting, the gap value is about $55\,$meV, similar to what has been reported previously \cite{Otrokov.nat2019, Estyunin.aplmat2020, Shikin.srep2020}. The second-derivative representation (Fig. \ref{ARPES}e) provides a clear illustration of the gapped Dirac cone. STS data recorded at 1.2 K (Fig. \ref{ARPES}f), i.e., well below the N\'eel temperature, show a $\text{d}I/\text{d}V$ spectrum featuring a local minimum near the expected DP position. This is compatible with the gap of the order of 50 meV in agreement with the result of the EDC analysis shown in Fig. \ref{ARPES}b. Even though the d$I$/d$V$ signal does not vanish at the (gapped) DP, the $\text{d}I/\text{d}V$ map at $-0.27\,$V displayed in Supplementary Fig.S4 is featureless and homogeneous, meaning that only a background signal is detected, i.e., no states are present. Similar featureless maps are observed in a range of  $-0.27\,\pm0.02$V, in line with the  DP gap of about 50 meV observed in ARPES (Fig. \ref{ARPES}a,b,e). For another sample, however, the laser-ARPES reveals a TSS with a substantially reduced gap, of the order of 20 meV (Fig. \ref{ARPES}c,d,g). Indeed, the $\overline \Gamma$ point EDC (Fig. \ref{ARPES}d) shows an apparent peak near the expected DP position. However, there is a shoulder at the high binding energy flank of the peak, which, according to the EDC fitting, is due to the lower part of the gapped TSS. The signatures of such a behavior are also seen in the STS (note that the samples studied in ARPES and STM/S are different, although they are from the same batch), showing a peak at -0.29 V (Fig. \ref{ARPES}h), consistent with the presence of the unresolved spectral features revealed by the EDC fitting at $\overline \Gamma$. In this case, as can be seen in Supplementary Fig. S5, the $\text{d}I/\text{d}V$ map at $-0.30\,$V exhibits a stronger signal, as a result of the contribution of the edges of the Dirac cone states observed also in the EDC (Fig. \ref{ARPES}d). The modulation of the $\text{d}I/\text{d}V$ signal is more affected in this case by the presence of the deep defects (see Supplementary Fig. S5c).  However, the overall shape of the spectra shown in Fig. \ref{ARPES}f and \ref{ARPES}h is independent of point defects (see Supplementary Fig. S4 and S5, respectively).

\section{Discussion}

We now discuss the possible origin of the DP gap size variation from sample to sample and within one sample. Well-defined dispersion lines observed with laser-ARPES indicate a reasonably good quality of the crystal surface, in agreement with our STM observations. 
Therefore, the surface SL crystal structure is largely similar to that of SLs in bulk and thus the near-surface magnetic structure should be the same as in bulk too, which has been recently confirmed using magnetic force microscopy \cite{Sass.prl2020}.
Thus, the variation (or a complete closing \cite{Hao.prx2019, Li.prx2019, Chen.prx2019, Swatek.prb2020, Nevola.prl2020, Yan.arxiv2021}) of the DP gap does not seem to come from a radical change of the magnetism at the surface. Neither it comes from some severe surface crystal structure degradation, not observed for the \MBT\, single crystals in ultra high vacuum. From the available STM and XRD evidence, apart from the unavoidable steps at the surface, the only significant structural imperfections of \MBT\, are related to point defects, caused by cation (Mn-Bi) intermixing.

The evidences presented here and in the literature indicate that these defects are formed in the bulk of the sample during its growth and then naturally find themselves near the surface because of the crystal cleavage before the ARPES or STM/S measurements.
The role of these defects for magnetic properties of \MBT\, and related compounds is being discussed currently \cite{Murakami.prb2019, Liu.prx2021, Lai.prb2021}. Recent high-field magnetization measurements show that reaching the saturation magnetization of \MBT\, (corresponding to about  4.6 $\mu_B$ per Mn) requires very large external magnetic fields of about 60 T \cite{Lai.prb2021}, while many previous studies revealed an incomplete saturation, $\sim$3 - 3.8 $\mu_B$ per Mn at about 6-7 T \cite{Otrokov.nat2019, Lee.prr2019, Yan.prm2019, Yan.prb2019, Li.prl2020, Li.pccp2020, Jiao.jsnm2021}. The reason for this has been found to be a "ferrimagnetic" structure of the septuple layer block, in which the local moments of the \mnbi defects are coupled antiparallel to those of the central Mn layer \cite{Riberolles.prb2021, Liu.APSMarch2021}. This is completely analogous to what is observed in \MSbT \cite{Liu.prx2021, Riberolles.prb2021}, which is a related isostructural compound \cite{Eremeev.jac2017, Eremeev.jpcl2021, Wimmer.advmat2021}. The essential difference between \MBT\, and \MSbT\, is a more pronounced cation intermixing in the latter \cite{Liu.prx2021, Riberolles.prb2021}, meaning a larger number of the Mn atoms at the Sb sites (Mn$_\text{Sb}$). Recent neutron diffraction measurements \cite{Liu.prx2021, Riberolles.prb2021} have shown the AFM coupling between the central Mn layer and the Mn$_\text{Sb}$ atoms. Due to a large amount of Mn$_\text{Sb}$ atoms in \MSbT\, the magnetic moment per Mn atom at 6-7 T is only about 2 $\mu_B$ \cite{Lai.prb2021}. Therefore, similarly to \MBT, very strong fields of up to 60-70 T are needed to overcome the \emph{intrablock} AFM coupling and fully polarize the SLs. In the Mn(Bi$_{1-x}$Sb$_x$)$_2$Te$_4$ solid solutions, the $M(H)$ behavior, observed in \MSbT, continuously evolves into that of \MBT \cite{Yan.prb2019, Lee.prx2021}. The latter facts strongly point towards the ferrimagnetic structure of the  \MBT\, SLs as well.

\begin{figure*}[!bth]
\begin{center}
\includegraphics[width=1\textwidth]{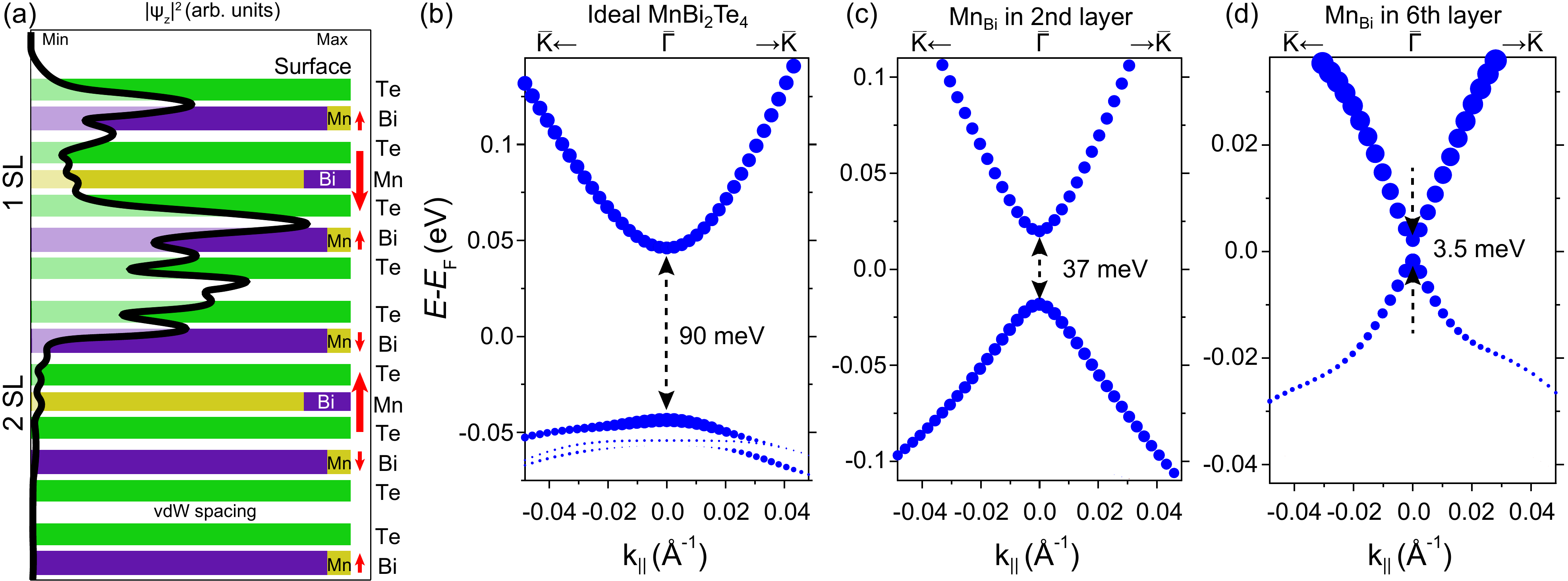}
\end{center}
\caption{(a) Illustration of the TSS real space distribution at the \MBT(0001) surface. The $|\Psi_z|^2$ profile corresponds to the band structure shown in panel (b). The color coding for the atoms sorts is the same as in Fig.\,\ref{fig:str}a. Unlike in the case of the ideal structure (Fig.\,\ref{fig:str}a), the Mn-Bi intermixing leads to appearance of the \mnbi magnetic moments in the Bi layers that are coupled antiparallel to those in the central Mn layer of the same SL. The \mnbi magnetic moments thus turn out to be located in the regions with a high weight of the TSS, strongly counteracting the effect from the magnetization of the central Mn layer, where the TSS weight is low. This is expected to lead to a strong reduction of the DP gap. 
The latter is illustrated in panels (b-d), where the \MBT\, surface electronic structure in the defectless case (b) is compared to those with Mn-Bi intermixing when \mnbi defect locates in the second (c) and sixth (d) atomic (Bi) layers counting from the surface (see panel (a)). Note that only the topological surface state is shown, while the bulk-like
bands are omitted. The energy axes scales in (b,c) and (d) are different.} \label{sketch}
\end{figure*}

The ferrimagnetic structure, along with the presence of \bimn\, in the Mn layer (as found by our structure refinement as well as in Refs. \cite{Zeugner.cm2019, Riberolles.prb2021, Ding.prb2020-2}), is expected to significantly reduce the effective magnetization of each individual SL block of \MBT. Eventually, at the surface, this should cause a decrease of the DP gap size. However, an approximately $20\,\%$ decrease of magnetization of each SL, which can be expected based on our XRD data, can hardly explain the DP gap size reduction by at least a factor of 2, as we observe in ARPES and STS. The reason why the cation intermixing should strongly affect the DP gap size becomes clear when the real space TSS distribution is analyzed. As it is shown in Fig.\,\ref{sketch}a, the weight of the TSS in the Te-Bi-Te trilayers of the surface SL is much larger than in the Te-Mn-Te trilayer. Thus, due to the Mn-Bi intermixing, the magnetization of \mnbi, counteracting the effect of the central layer Mn atoms, is introduced exactly in the regions of the TSS predominant localization. In turn, the central Mn layer, where the TSS weight is small, becomes slightly “magnetization depleted” due to the Bi atoms incorporation. Depending on the intermixing levels the cooperation of these two factors may result in a significant reduction of the size of the DP gap or even in its almost complete shrinking, a phenomenon that has been observed \cite{Hao.prx2019, Li.prx2019, Chen.prx2019, Swatek.prb2020, Nevola.prl2020, Yan.arxiv2021}, but not satisfactorily explained up to now. 

To confirm the above suggested scenario of the DP gap reduction, we have performed fully-relativistic DFT surface electronic structure calculations of \MBT(0001) (see Methods section for the calculation details). It can be seen in Fig. \ref{sketch}b that the pristine \MBT\, surface features the DP gap of 90 meV, in agreement with previous calculations \cite{Otrokov.nat2019, Otrokov.prl2019}. Then, when a pair of Mn and Bi atoms are exchanged so that the \mnbi\, atom goes in the subsurface atomic layer, the DP gap appears to be reduced by about 2.5 times, i.e. to 37 meV (Fig. \ref{sketch}c). This already shows that the antiparallel alignment of the magnetic moments of \mnbi\, (with respect to the central Mn layer) has an important effect on the \MBT\, DP gap size. However, while the topological surface state charge density shows a local maximum around the subsurface Bi layer, the second Bi layer (i.e. the 6th atomic layer counting from the surface) carries much larger weight of the state (Fig. \ref{sketch}a). Remarkably, introducing \mnbi\, defect in the 6th layer leads to almost complete shrinking of the DP gap (Fig. \ref{sketch}d), whose calculated value amounts to 3.5 meV, i.e. by about 25 times smaller than in the defectless case. These results provide a theoretical proof that the \mnbi\, defects can cause a strong reduction of the \MBT\, DP gap due to the ferrimagnetic structure of the SL \cite{Liu.prx2021, Riberolles.prb2021, Lai.prb2021} and predominant localization of the topological surface state in the Bi layers of the surface SL block.

A recent study reports\cite{Liu.prx2021} that the degree of the cation intermixing in \MSbT\, may be varied by changing the growth temperature. In \MBT, the Mn-Bi intermixing should be sensitive to the growth temperatures and starting compositions too and, therefore, one can expect that the degree of it  may differ from sample to sample or may even experience certain variations in the same crystal. Indeed, the per layer concentrations of the \mnbi atoms in the single crystal samples by different groups are reported to range from 2.5 to 5 \% (4.6-5.7 \%) as estimated based on the STM measurements \cite{Yan.prm2019, Yuan.nl2020, Liang.prb2020, Huang.prm2020} (XRD measurements in this work and Ref. \cite{Zeugner.cm2019}). Besides, our STM measurements indicate the fluctuation of the \mnbi concentration within one sample (Supplementary Fig. S6). It seems like such variations of the defects concentrations may also affect the free carriers concentration and the value of the Néel temperature  \cite{Otrokov.nat2019, Lee.prr2019, Zeugner.cm2019, Yan.prm2019}, that slightly varies from sample to sample within the 24-25.4 K range (Supplementary Fig. 9 and the corresponding note). These arguments provide a plausible explanation of the reason of the observation of the different DP gap values in different samples.

Let us now discuss the here proposed DP gap reduction scenario in context of the temperature induced transition into the paramagnetic phase. When the Dirac cone is nearly gapless, then no strong changes are expected to be seen in ARPES upon heating above the N\'eel temperature. However, when the DP gap is sizable, the magnetic nature of the here proposed mechanism implies that it should close in the high-temperature magnetically disordered state. While the early synchrotron ARPES studies reported \cite{Otrokov.nat2019, Vidal.prb2019, Lee.prr2019} that the DP gap in \MBT\, persists well above the N\'eel temperature, recent laser-ARPES data show about 40\% reduction of the DP gap (from 65 to 40 meV) upon heating from below $T_N$ up to 35 K \cite{Shikin.arxiv2021}. An incomplete closing of the gap seems to be consistent with strong short range order effects that persist in \MBT\, up to about 50-60 K, as observed by electron spin resonance, ferromagnetic resonance, and antiferromagnetic resonance experiments \cite{Otrokov.nat2019, Alfonsov.prb2021, Alfonsov.arxiv2021}. The measured magnetization data \cite{Lai.prb2021}, revealing that \MBT\, is not in the paramagnetic limit even at $T \approx 50$ K, confirm this observation. Such a behavior is also consistent with the strong spin fluctuation-driven spin scattering above $T_N$ found in a previous magneto-transport study of \MBT\, in Ref. [\onlinecite{Lee.prr2019}]. 
Beyond 50-60 K, an unprecedentedly large anisotropy of the Mn spin relaxation rate in the paramagnetic state of \MBT\, \cite{Otrokov.nat2019, Alfonsov.prb2021} may give rise to an instantaneous (on the timescale of electron spin resonance) out-of-plane magnetic field at the surface, preventing the gap to close even at $T > T_N$ on the \emph{much faster timescale of the ARPES experiment}. 

The latter interpretation of the ARPES data imply that the DP gap closing should in principle be observable with other techniques. Recently, local measurements with point-contact tunneling spectroscopy have allowed a detection of the magnetic gap at the DP of \MBT\, at some surface locations \cite{Ji.arxiv2019}. Although in other surface areas there was no gap detected by the same technique, this does not contradict the here proposed scenario based on the crucial role of the Mn-Bi intermixing, since the degree of the intermixing may vary across the surface. Indeed, as we have written above, our STM measurements show that the \mnbi\, concentration fluctuates across the surface at the $100 \times 100$ nm$^2$ scale, at least judging by the \mnbi\, concentration in the subsurface layers. We have also found that cleaving the sample exposes a new surface with the same property, i.e. a fluctuating concentration of \mnbi across the surface. Thus, a larger (smaller) average concentration of \mnbi in XRD (or any other integral technique) will not straightforwardly translate into a smaller (larger) DP gap in STS: the size of the latter will be a rather local property as compared to the scale of the sample size. Moreover, it might well be a local property even in $\mu$-laser-ARPES, although at a larger scale (light spot is about 5 $\mu$m). Indeed, the ARPES mapping of the \MBT\, surface shows that the electronic structure is inhomogeneous on the scale of 100 - 150 $\mu$m (see Supplementary Fig. 4 of Ref. [\onlinecite{Estyunin.aplmat2020}]). Ideally, an \emph{in situ} study of the very same surface and its very same local area by low-temperature $\mu$-laser ARPES and low-temperature STM/S in the same instrumental setup is required. However, given a highly different spatial scales of the ARPES and STM, this appears to be hardly feasible. Our results thus highlight a necessity to suppress the cation (Mn-Bi) intermixing in \MBT\, thus minimizing the number of \mnbi,  which is a crucial task for the nearest future studies. Improving the structural quality of \MBT\, up to the level of the state-of-the-art samples of its parent compound \BT\, \cite{Sessi.prb2013, Netsou.acsn2020} will hopefully allow getting rid of the DP gap issue in this AFM TI.

\section{Conclusions}

In conclusion, we have experimentally studied the crystalline and electronic structure of the (0001) surface of the AFM TI \MBT\, using STM/S, micro($\mu$)-laser ARPES, and first-principles calculations. On a large scale, the surface appears to be atomically flat, with several hundreds nanometer wide terraces, separated by septuple layer high steps. On the atomic scale, a well-defined hexagonal lattice is detected with a $(1\times 1)$ periodicity, i.e., no reconstruction or degradation occurs upon cleavage, as generally expected for a 2D van der Waals material. Further, we clearly observe two kinds of spectroscopic features on the topographic STM images. Namely, we distinguish (i) circular protrusions stemming from the Bi atoms at the surface Te sites and (ii) triangular depressions due to the Mn atoms at the subsurface Bi sites. The presence of the Mn atoms in the subsurface Bi layer indicates that they are located in the sixth Bi layer, too, while Bi atoms, in turn, occupy Mn positions in the fourth layer (cation intermixing), which is strongly supported by the results of our X-ray diffraction data refinement.

Our low temperature STS/$\mu$-laser-ARPES experiments reveal that the size of the Dirac point gap in the topological surface state differs for different cleavages/samples. We attribute this behavior to the effect of the spatially inhomogeneous cation (Mn-Bi) intermixing, which affects the distribution of Mn atoms and, according to the recent high-field magnetization measurements \cite{Lai.prb2021}, leads to a deviation of the \MBT\, septuple layer magnetic structure from the ideal ferromagnetic to a \emph{ferri}magnetic one. The latter structure along with the predominant real-space localization of the topological surface state around the Bi layers of the topmost septuple layer lead to a dramatic reduction of the Dirac point gap size, as revealed by our first-principles electronic structure calculations. A variation of the degree of defectness should lead to a different exchange splitting of the Dirac point for different samples or sample cleavages. 

\section*{Methods}

\subsection*{Crystal growth}

The bulk \MBT\, single crystals were grown by the modified Bridgman method. To perform a careful refinement of the \MBT\, crystal structure, we synthesized polycrystalline single-phase samples, that contained no MnTe or any other phases. The synthesis was carried out in sealed quartz ampoules by melting elements taken in stoichiometric ratios. Samples of the polycrystalline alloy were ground, pressed into pellets, and annealed at 575$^\circ$C. This process was repeated three times with a total annealing time of 750 hours. The diffraction pattern obtained and the corresponding structure refinement are presented in the Supplementary Information.

\subsection*{STM/S measurements}

STM/S measurements were performed on a custom-designed ultra high vacuum (UHV) system equipped with a low temperature scanning tunneling microscope. The crystal was cleaved by Nitto tape \textit{in situ} at room temperature and directly transferred to the STM. The base pressure during the experiments was $ 2 \rm\times 10^{-10}$ mbar. STM images were recorded in constant current mode and the differential conductance ($\text{d}I/\text{d}V$) spectra were taken using a lock-in amplifier (f = 763.7 Hz) at $4.7\,$K and $1.2\,$K. The images were processed using the WSxM software \cite{Horcas.WsXM2007}. 

\subsection*{ARPES measurementes}

The ARPES measurements were carried using $\mu$-laser with improved angle and energy resolution and a laser beam with a spot diameter around 5 $\mu$m, using a Scienta R4000 electron energy analyzer with an incidence angle of 50$^{\circ}$ relative to the surface normal. The measurements were performed using $p$-polarized laser radiation with a photon energy $h\nu=6.3$ eV at a temperature of 10 K.

\subsection*{Resistivity measurements}

Resistivity measurements were done with a standard four-probe ac technique using a low-frequency (f$ \approx$ 20\,Hz) lock-in amplifier. Contacts were attached with conducting graphite paste.

\subsection*{DFT calculations}
Electronic structure calculations were carried out within the density functional theory using the projector augmented-wave (PAW) method \cite{Blochl.prb1994} as implemented in the VASP code \cite{vasp1,vasp2}. The exchange-correlation energy was treated using the generalized gradient approximation \cite{Perdew.prl1996}. The energy cutoff for the plane-wave expansion was set to 270 eV. The Mn $3d$-states were treated employing the GGA$+U$ approach \cite{Anisimov1991} within the Dudarev scheme \cite{Dudarev.prb1998}. The $U_\text{eff}=U-J$ value for the Mn 3$d$-states was chosen to be equal to 5.34~eV, as in previous works \cite{Otrokov.jetpl2017, Otrokov.2dmat2017,Eremeev.jac2017, Hirahara.nl2017, Eremeev.nl2018, Otrokov.prl2019, Otrokov.nat2019, Wimmer.advmat2021, Jahangirli.jvstb2019, Petrov.prb2021}.

STM/S simulations were performed using the Tersoff-Hamann approximation. We have chosen a ($5 \times 3\sqrt{3}$) rectangular cell (21.68 \AA\, $\times$ 22.53 \AA; about 420 atoms), containing two \MBT\ septuple layers and a vacuum layer with a thickness of no less than 10 \AA. Structural optimizations were performed using a conjugate-gradient algorithm and a force tolerance criterion for convergence of 0.01 eV/{\AA}. The $\overline \Gamma$-centered $k$-point meshes of $2\times2\times1$ and $3\times3\times1$ were used to sample the 2D Brillouin zone for the relaxations and static calculations, respectively. In order to describe the van der Waals interactions we made use of the DFT-D3 \cite{Grimme.jcp2010, Grimme.jcc2011} approach. Spin-orbit coupling was neglected.

In the surface electronic structure calculations, the Hamiltonian contained scalar relativistic corrections and the spin-orbit coupling was taken into account by the second variation method \cite{Koelling.jpc1977}. The 6-SL-thick slab and the (3 $\times$ 3) in-plane supercell have been chosen (378 atoms). The $\overline \Gamma$-centered $k$-point mesh of $3\times3\times1$ has been used. We have compared three following cases: (i) ideal \MBT, and \MBT\, with one \mnbi\, substitution in the (ii) 2nd and (iii) 6th atomic layers ($\sim$11 \% per layer). In the latter two cases, the substituted Bi atom has been placed in the central (Mn) layer of the surface SL. The defects have only been introduced in the surface SL because the topological surface state charge density is largest there (see Fig. \ref{sketch}a). Structural relaxations due to the introduction of the Mn-Bi intermixing were neglected. The ferrimagnetic structure of the surface SL has been assumed in which the local moment of the \mnbi\, defect is coupled antiparallel to those of the central Mn layer \cite{Lai.prb2021,Liu.prx2021, Riberolles.prb2021}.


\bibliographystyle{apsrev}

\section*{Acknowledgments}

The authors thank M. Ilyn, M.A. Valbuena and S.V. Eremeev for stimulating discussions. We acknowledge support by the Spanish Ministerio de Ciencia e Innovacion (Grant no. PID2019-103910GB-I00, PGC2018--093291--B--I00, PGC2018-097028--A--I00 and PGC2018-098613--B--C21) and Saint Petersburg State University (project ID No. 73028629). IMDEA Nanociencia acknowledges support from the 'Severo Ochoa' Programme for Centres of Excellence in R$\&$D (MINECO, Grant SEV-2016-0686). M.G. has received financial support through the Postdoctoral Junior Leader Fellowship Programme from “la Caixa” Banking Foundation. Z. A. and  N.M. acknowledge the support of the Science Development Foundation under the President of the Republic of Azerbaijan (Grant No. EIF-BGM-4-RFTF-1/2017-21/04/1-M-02). I.I.K., D.E. and A.M.S. acknowledge the support from  Russian Science Foundation (Grant No. 18-12-00062) and Russian Foundation of Basic Researches (Grant No. 20-32-70179).






\end{document}